\documentclass{article}

\usepackage{PRIMEarxiv}

\usepackage[utf8]{inputenc} 
\usepackage[T1]{fontenc}    
\usepackage{hyperref}       
\usepackage{url}            
\usepackage{booktabs}       
\usepackage{amsfonts}       
\usepackage{nicefrac}       
\usepackage{microtype}      
\usepackage{lipsum}
\usepackage{fancyhdr}       
\usepackage{graphicx}       
\usepackage{xcolor}
\usepackage{diagbox} 

\usepackage{caption}
\usepackage{subcaption}

\usepackage{amsmath,amssymb} 
\usepackage{algorithm}
\usepackage{algorithmic} 
\usepackage{multirow}
\usepackage{makecell}
\usepackage{tabularray}

\usepackage[textsize=scriptsize]{todonotes}\reversemarginpar

\graphicspath{{media/}}     

\title{On Jailbreaking Quantized Language Models Through Fault Injection Attacks}

\author{Noureldin Zahran$^\dagger$, Ahmad Tahmasivand$^\ddagger$, Ihsen Alouani$^\S$, Khaled Khasawneh$^\ddagger$ and Mohammed E. Fouda$^\dagger$\\
  $^\dagger$Compumacy for Artificial Intelligence Solutions, Cairo, Egypt \\
  $^\ddagger$Electrical and Computer Engineering Department, George Mason University, VA, USA\\
  $^\S$School of Electronics, Electrical Engineering and Computer Sciences, Queen's University Belfast, United Kingdom\\
  \texttt{Emails: Nour@compumacy.com, atahmasi@gmu.edu, i.alouani@qub.ac.uk,}\\\texttt{kkhasawn@gmu.edu, foudam@uci.edu}   
}

\begin{document}
\maketitle

\begin{abstract}
The safety alignment of Language Models (LMs) is a critical concern, yet their integrity can be challenged by direct parameter manipulation attacks, such as those potentially induced by fault injection. As LMs are increasingly deployed using low-precision quantization for efficiency, this paper investigates the efficacy of such attacks for jailbreaking aligned LMs across different quantization schemes. We propose gradient-guided attacks, including a tailored progressive bit-level search algorithm introduced herein and a comparative word-level (single weight update) attack. Our evaluation on Llama-3.2-3B, Phi-4-mini, and Llama-3-8B across FP16 (baseline), and weight-only quantization (FP8, INT8, INT4) reveals that quantization significantly influences attack success. While attacks readily achieve high success (>80\% Attack Success Rate, ASR) on FP16 models, within an attack budget of 25 perturbations, FP8 and INT8 models exhibit ASRs below 20\% and 50\%, respectively. Increasing the perturbation budget up to 150 bit-flips, FP8 models maintained ASR below 65\%, demonstrating some resilience compared to INT8 and INT4 models that have high ASR. In addition, analysis of perturbation locations revealed differing architectural targets across quantization schemes, with (FP16, INT4) and (INT8, FP8) showing similar characteristics. Besides, jailbreaks induced in FP16 models were highly transferable to subsequent FP8/INT8 quantization (<5\% ASR difference), though INT4 significantly reduced transferred ASR (avg. 35\% drop). These findings highlight that while common quantization schemes, particularly FP8, increase the difficulty of direct parameter manipulation jailbreaks, vulnerabilities can still persist, especially through post-attack quantization.
\end{abstract}


\section{Introduction}
\label{sec:intro}

Recent advancements in LMs have led to their widespread integration into diverse applications, fundamentally changing how users interact with information and technology. Increasingly, LMs serve as conversational assistants, sometimes supplanting traditional search engines as primary sources for answers and information generation \cite{consq_genai, AlixPartners2024FutureSearch}. This central role necessitates a strong focus on model safety and alignment, ensuring that LM-generated responses are helpful, and harmless, aligning with human values and preventing their misuse for generating malicious or unethical content. Consequently, significant effort has been invested in aligning these models through techniques such as Supervised Fine-Tuning and Reinforcement Learning from Human Feedback \cite{wang2024comprehensivesurveyllmalignment}.

Despite these alignment efforts, LMs remain vulnerable to \emph{jailbreak} attacks, which aim to circumvent safety guardrails and elicit prohibited outputs. Most of the literature has focused on prompt-based attacks, where adversaries craft malicious inputs through methods like prompt engineering or adversarial tokens' optimization (e.g., GCG) to bypass safety protocols \cite{jb_survey}. Another vector involves adversarial fine-tuning, where even small amounts of targeted data can degrade a model’s safety alignment. These attacks, however, typically operate at the software or input level \cite{unlearning_fail}.

This paper focuses on an emerging class of vulnerability with a distinct attack vector, i.e., jailbreak attacks using bit flips within the model’s parameters. Bit-flip attacks (BFAs) represent a fundamentally different threat vector, originating from hardware-level phenomena rather than input manipulation. Bit flips can occur naturally due to environmental factors such as cosmic rays, electromagnetic interference, or voltage faults \cite{oles2024understanding, li2021securing}. Critically, they can also be induced maliciously through physical fault-injection techniques such as Rowhammer \cite{kim2020revisiting, OG_ROWHAM}. The Rowhammer attack, in particular, exploits DRAM vulnerabilities where repeatedly accessing specific memory rows (\emph{aggressor rows}) causes charge leakage in adjacent rows, leading to unintended bit flips ($0\!\rightarrow\!1$ or $1\!\rightarrow\!0$) in victim rows. This allows an adversary, potentially with only user-level privileges, to surgically alter stored data, including the parameters of a deployed LM.

Prior work extensively studied BFAs targeting convolutional neural networks (CNNs); researchers demonstrated that flipping critical bits could cause catastrophic performance degradation \cite{OGBFA}, targeted misclassification \cite{TBFA}, or enable back-door injection \cite{proflip}, often assuming white-box access to model parameters to identify vulnerable bits using optimization-based search algorithms. Recently, Coalson \emph{et al.} introduced \textit{PrisonBreak} \cite{prisonbreak}, the first work that shows that BFAs can effectively jailbreak aligned LMs. Flipping only 25 bits in the model weights can bypass safety alignment and cause models to generate persistently harmful content, without requiring any malicious user prompt. This established BFA jailbreaking as a pressing and practical threat.

However, the \textit{PrisonBreak} study focused primarily on LMs using half-precision (FP16) parameters. Current trends in LM deployment increasingly favor quantization, utilizing lower-precision formats such as FP8 or INT8 for weights and/or activations. Quantization significantly reduces the memory footprint and computational cost, enabling efficient inference on resource-constrained hardware, including edge devices \cite{quant_survey}. This shift towards quantization raises a critical question: \emph{How effective are BFA jailbreak techniques against quantized LMs?} Insights from earlier BFA work on CNNs suggest that quantization might increase robustness; attacking full-precision (FP32) models can be simpler by targeting exponent bits, causing large-magnitude changes, whereas quantization formats lack these exponents and are inherently more robust to random flips,  and their discretized value steps can further hinder gradient-based search algorithms, necessitating more sophisticated targeted bit-search algorithms \cite{OGBFA}. We hypothesise that this effect extends to LMs, potentially making quantized models harder to jailbreak via BFAs compared to their FP16 counterparts. The key contributions are summarized as follows:

\begin{itemize}
\item \textbf{Evaluation of BFA Jailbreaking on Quantized LMs using a Tailored Algorithm.} We systematically assess the feasibility and effectiveness of BFA jailbreaks in various popular quantization formats. This evaluation utilizes a tailored greedy, gradient-based search algorithm, introduced herein, specifically designed to identify critical bits for jailbreaking within the context of quantized LMs.
\item \textbf{Comparison of Attack Strategies on Modern LMs.} We investigate different BFA strategies by comparing progressive bit-level and word-level attacks. This comparison is conducted on newer, publicly available aligned LMs, including \textit{Phi-4-mini} and \textit{Llama 3.2}.
\end{itemize}

The remainder of this paper is structured as follows: Section~\ref{sec:related_work} provides background on previous bit-flip attacks. Section ~\ref{sec:threat_model} defines the threat model. Section~\ref{sec:methodology} details the attack objective, the attack algorithms, and the quantization handling approach. Section ~\ref{sec:exp_setup} states the selected models and parameters, dataset creation, and the evaluation strategy. Section~\ref{sec:results} presents the results and Section~\ref{sec:conclusion} concludes the paper.

\section{Related Work}
\label{sec:related_work}
BFAs represent a class of hardware-level threats targeting the parameters of neural networks. These attacks leverage physical phenomena to alter the binary representation of model weights stored in memory, potentially leading to significant changes in model behavior. The execution of a BFA typically involves two phases: an offline preparation stage and an online execution stage. In the offline stage, the attacker, often assumed to have access to a replica of the target model (white-box assumption), performs software simulations to identify the most critical bit locations whose alteration maximally impacts a chosen objective function. This preparatory step is crucial for optimizing the attack and minimizing the interaction required with the actual hardware during the online phase, where techniques like Rowhammer might be employed to physically induce the calculated bit flips.

Identifying the optimal set of bits to flip is computationally challenging due to the vast parameter space. Early work on BFAs targeting CNNs developed methodologies to address this. Rakin et al \cite{OGBFA} introduced a greedy, gradient-ascention based approach. The core idea was to use the gradient of a loss function (e.g., classification error) with respect to the model weights as a signal to identify potentially impactful bits. The BFA algorithm implemented a layer-wise search strategy. Within each layer, bits were scored based on their gradient magnitude, weighted by the change in weight value (`step size') caused by flipping that specific bit. Candidate bits with the highest scores were selected within a defined computational budget (e.g., the top 100 scoring bits). Each candidate bit was then individually evaluated via simulation: the bit was flipped, the model's loss was computed, and the bit was reverted. This allowed identifying the single most effective bit flip within the budget for that layer. The attack proceeded progressively: the globally most impactful bit identified across all layers in an iteration was permanently flipped, and subsequent iterations recalculated gradients based on the altered model state. This progressive, layer-fair approach was shown to be highly effective, capable of inducing near-random classification accuracy in models like ResNet-18 with only a small number (e.g., 13) of targeted bit flips, significantly outperforming random bit corruption. Importantly, the BFA paper demonstrated the attack's efficacy even against quantized networks (4, 6, and 8-bit integer representations), highlighting that targeted search could overcome the presumed increased robustness of lower-precision models compared to random perturbations. The primary objective this study was model destruction or severe performance degradation.

While Bitflip-based attacks were established as a potent threat for CNNs, their specific application to LMs for the purpose of jailbreaking is a more recent area of investigation. PrisonBreak is the first to demonstrate the feasibility of this specific threat. Moving beyond the goal of simple performance degradation, PrisonBreak aimed to precisely un-align LMs, forcing them to comply with harmful instructions they were trained to refuse, while ideally maintaining their general utility on benign tasks. This research targeted contemporary, aligned LMs such as Llama 2,3 and Vicuna, deployed using half-precision floating-point (FP16) parameters.

Adapting BFA techniques to the generative nature and safety alignment of LMs required several key innovations\cite{prisonbreak}. A major challenge was that simply optimizing the model to start its response affirmatively (e.g., "Sure, here is...") which was enough for earlier prompt-based gradient attacks\cite{GCG} often led to degenerate outputs like overfitting to the affirmative phrase or nonsensical repetition. To address this and ensure the generation of coherent, harmful content, PrisonBreak proposed using a \textit{proxy dataset} of full harmful completions (generated using existing uncensored models) as optimization targets. Furthermore, recognizing the causal nature of autoregressive generation, where initial tokens heavily influence subsequent ones, they introduced a \textit{modified cross-entropy loss function}. This loss incorporated exponentially decaying weights across the token sequence, prioritizing the accurate generation of the initial parts of the target response (both the affirmative opening and the beginning of the harmful content). To manage the computational expense of searching parameters in billion-scale LMs, PrisonBreak also implemented several search space reductions based on empirical findings: the search was primarily limited to the three most significant exponent bits in the FP16 representation, focused on 0$\rightarrow$1 flips, and excluded layers less likely to be impactful for their objective (embedding, unembedding, layer normalization). Leveraging these contributions, PrisonBreak demonstrated high jailbreak success rates across multiple LMs using fewer than 25 targeted bit flips, achieved without any modification to the user's input prompt.

\section{Threat Model}
\label{sec:threat_model}

Similar to previous work \cite{prisonbreak, OGBFA, proflip, TBFA}, we consider an adversary whose objective is to jailbreak LMs through direct manipulation of their parameters stored in memory using bit-flip attacks. The assumed operational context involves LMs deployed in shared computing environments, such as MLaaS platforms, where the service provider attempts to maintain safety alignment. We adopt a white-box attacker model, granting the adversary knowledge of the target model's architecture and parameters, a standard assumption in related BFA literature and a practical consideration given the rise of open-source LMs. However, the attacker does not possess or require access to the original training data used for the LM. The attack requires the adversary to co-locate a process or virtual machine (VM) on the same physical host as the victim's LM instance, operating with only standard user-level privileges.

\section{Methodology}
\label{sec:methodology}
This section details the core approach used for inducing jailbreaks in LMs via parameter manipulation. In this section, we define the attack objective, describe the algorithms used to achieve it, and explain the necessary techniques for applying these algorithms to quantized models.

\subsection{Attack Objective}
\label{ss:atk_obj}
The primary objective of our parameter manipulation attack is to induce jailbreak (JB) behavior. This involves modifying the LM's parameters, $\theta$, such that it generates harmful or undesirable content in response to specific queries, thereby bypassing its safety alignment. The general goal is to minimize a chosen loss function, $\mathcal{L}_{JB}$, which quantifies the difference between the model's output and a target harmful completion, given the corresponding harmful query.

The attack algorithms, detailed in the next section, achieve this by modifying the model parameters iteratively. Starting with the initial parameters $\theta^{(0)} = \theta_{orig}$, the parameters at iteration $i$ are determined by applying the single best perturbation found during that iteration to the parameters from the previous iteration, $\theta^{(i-1)}$. Specifically, let $\mathcal{P}_i(\theta^{(i-1)})$ be the set of candidate parameter states reachable from $\theta^{(i-1)}$ by applying one valid perturbation (a single bit flip or a single weight update) identified by the attack algorithm in iteration $i$. The parameters for the next iteration are chosen greedily to minimize the objective function:
\begin{equation} \label{eq:iter_min_obj}
\theta^{(i)} = \underset{\theta' \in \mathcal{P}_i(\theta^{(i-1)})}{\operatorname{argmin}} \mathcal{L}_{JB}(\theta') \quad \text{for } i = 1, \dots, N_{iter}
\end{equation}
This iterative process continues for a predefined number of steps, $N_{iter}$.

Following the approach established by Coalson et al. \cite{prisonbreak}, our specific instantiation of $\mathcal{L}_{JB}$ is designed not merely to elicit an initial affirmative response (e.g., "Sure, here is...") but to maximize the likelihood of the model generating a more complete, harmful response sequence. This strategy aims to mitigate issues observed in prior attack attempts where targeting only affirmative prefixes led to repetitive or nonsensical outputs. For this purpose, we adopt the jailbreaking objective function proposed by Coalson et al. \cite{prisonbreak}, defined as:
\begin{equation} \label{eq:jailbreak_loss}
\mathcal{L}_{JB}(\theta) = - \frac{1}{n} \sum_{i=1}^{n} \sum_{k=1}^{m} e^{-\left(\frac{k-1}{m-1}\right)} \log f_{\theta}(y_k | s_{k-1})
\end{equation}
where $(x, y)$ represents a prompt-target pair from the dataset used to guide the attack, $n$ is the dataset size, $s_{k-1} = (x_1, ..., x_n, y_1, ..., y_{k-1})$ are the tokens up to position $k-1$, $f_{\theta}(y_k|s_{k-1})$ is the model's predicted probability for token $y_k$, and $m$ is the total number of tokens in the target response $y$. This loss function applies exponentially decaying weights to the standard cross-entropy loss across the target token sequence, thereby focusing more of the optimization effort on accurately generating the initial parts of the harmful completion.

\subsection{Attack Algorithms}
We employ greedy, gradient-based search strategies to identify and apply minimal parameter perturbations that achieve the jailbreaking objective defined previously. We implement two primary algorithms: a precise bit-level attack and a comparative word-level (weight) attack.

\textbf{Bit-Level Attack.}
The first algorithm, outlined in Algorithm~\ref{alg:bit_flip_main}, iteratively identifies and flips a single bit within the model's parameters per iteration to minimize the jailbreak loss $\mathcal{L}_{JB}$. This approach builds upon the progressive search methodology introduced in prior work \cite{OGBFA} but incorporates a modified strategy for selecting the specific bit to flip within candidate weights.


\begin{algorithm}[!t] 
\caption{Progressive Bit-Flip Attack (Main Loop)}
\label{alg:bit_flip_main}
\begin{algorithmic}[1]
\REQUIRE Language Model $M$, Attack Dataset $D_{atk}$, Loss $\mathcal{L}_{JB}$, $N_{iter}$, $N_{CL}$, $E_{max}$
\ENSURE Jailbroken Language Model $M$

\FOR{$i = 1$ \TO $N_{iter}$}
    \STATE \texttt{// --- Gradient Calculation ---}
    \STATE Zero gradients; Enable gradients; $L_{current} = \mathcal{L}_{JB}(M, D_{atk})$; Compute gradients $G$; Disable gradients.

    \STATE \texttt{// --- Candidate Evaluation ---}
    \STATE $C_{global\_best} \leftarrow \text{None}$; $L_{global\_best} \leftarrow L_{current}$.

    \FOR{each relevant layer $l$ in $M$}
        \STATE $e_{count} \leftarrow 0$; $W_l, G_l \leftarrow$ GetWeightsAndGrads($l$);
        \STATE $Idx_{W} \leftarrow$ Top $N_{CL}$ weight indices in $l$ by $|G_l|$; Sort $Idx_{W}$.
        \STATE $C_{layer\_best} \leftarrow \text{None}$; $L_{layer\_best} \leftarrow L_{current}$.

        \FOR{each candidate weight index $idx_w \in Idx_{W}$}
            \IF{$e_{count} \ge E_{max}$}
                \STATE \textbf{break} 
            \ENDIF
            \STATE $w_c = W_l[idx_w]$; $g_c = G_l[idx_w]$;
            \STATE $S_{bits} \leftarrow$ CalculateBitScores($w_c, g_c$);
            \STATE $Idx_{B} \leftarrow$ Sort bit indices by $S_{bits}$ (descending).

            \STATE $(j_{best}, L_{best}, e_{count}) \leftarrow FindBestBitInWeight(\dots)$

            \IF{$j_{best}$ is not None and $L_{best} < L_{layer\_best}$}
                \STATE $L_{layer\_best} \leftarrow L_{best}$.
                \STATE $C_{layer\_best} \leftarrow (l, idx_w, j_{best}, L_{best})$.
            \ENDIF
        \ENDFOR \COMMENT{End weight loop}

        \IF{$C_{layer\_best}$ is not None and $L_{layer\_best} < L_{global\_best}$}
             \STATE $L_{global\_best} \leftarrow L_{layer\_best}$.
             \STATE $C_{global\_best} \leftarrow C_{layer\_best}$.
        \ENDIF
    \ENDFOR \COMMENT{End layer loop}

    \STATE \texttt{// --- Apply Best Flip ---}
    \IF{$C_{global\_best}$ is None}
        \STATE Print "Early stopping."; \textbf{break}
    \ENDIF
    \STATE Permanently flip bit specified by $C_{global\_best}$ in $M$.
\ENDFOR \COMMENT{End main iteration loop}

\RETURN Modified Model $M$.
\end{algorithmic}
\end{algorithm}

At each iteration $i$, the algorithm first computes the gradients $G$ of the current loss $L_{current}$ with respect to the model weights. Within each relevant layer $l$, it identifies the top $N_{CL}$ weight parameters with the highest gradient magnitudes $|G_l|$ as candidates for modification. For each candidate weight $w_c$ (with index $idx_w$), the algorithm proceeds to find the most effective single bit flip using the subroutine \textit{FindBestBitInWeight} (detailed in Algorithm~\ref{alg:find_best_bit}).


\begin{algorithm}[h] 
\caption{FindBestBitInWeight}
\label{alg:find_best_bit}
\begin{algorithmic}[1]
\REQUIRE Model $M$, Dataset $D_{atk}$, Loss $\mathcal{L}_{JB}$, Layer $l$, Weight index $idx_w$, Original weight $w_c$, Sorted bit indices $Idx_{B}$, Current eval count $e_{count}$, Max eval count $E_{max}$
\ENSURE Best bit index $j_{best\_bit}$, Best loss $L_{best\_bit}$, Updated eval count $e_{count}$

\STATE $j_{best\_bit} \leftarrow \text{None}$
\STATE $L_{best\_bit} \leftarrow$ $\infty$

\FOR{each bit index $j \in Idx_{B}$}
    \IF{$e_{count} \ge E_{max}$}
        \STATE \textbf{break} \COMMENT{Layer evaluation budget exceeded}
    \ENDIF

    \STATE Create temporary flipped weight $w'_c$ by flipping bit $j$ in $w_c$.
    \STATE Temporarily update layer $l$ in $M$ with $w'_c$.
    \STATE Calculate loss $L'_{j} = \mathcal{L}_{JB}(M, D_{atk})$. \COMMENT{Loss evaluation}
    \STATE $e_{count} \leftarrow e_{count} + 1$.
    \STATE Revert model change (restore original $w_c$).

    \IF{$L'_{j}$ is invalid (NaN/Inf)}
        \STATE \textbf{continue}
    \ENDIF \COMMENT{Skip invalid results}
    \IF{$L'_{j} < L_{best\_bit}$}
        \STATE $L_{best\_bit} \leftarrow L'_{j}$.
        \STATE $j_{best\_bit} \leftarrow j$.
    \ELSIF{$L'_{j} \ge L_{best\_bit}$}
         \STATE \textbf{break} \COMMENT{Stop searching bits for this weight}
    \ENDIF
\ENDFOR \COMMENT{End bit loop}

\RETURN $j_{best\_bit}, L_{best\_bit}, e_{count}$
\end{algorithmic}
\end{algorithm}

\begin{figure}[h] 
    \centering 
    \includegraphics[width=0.8\columnwidth]{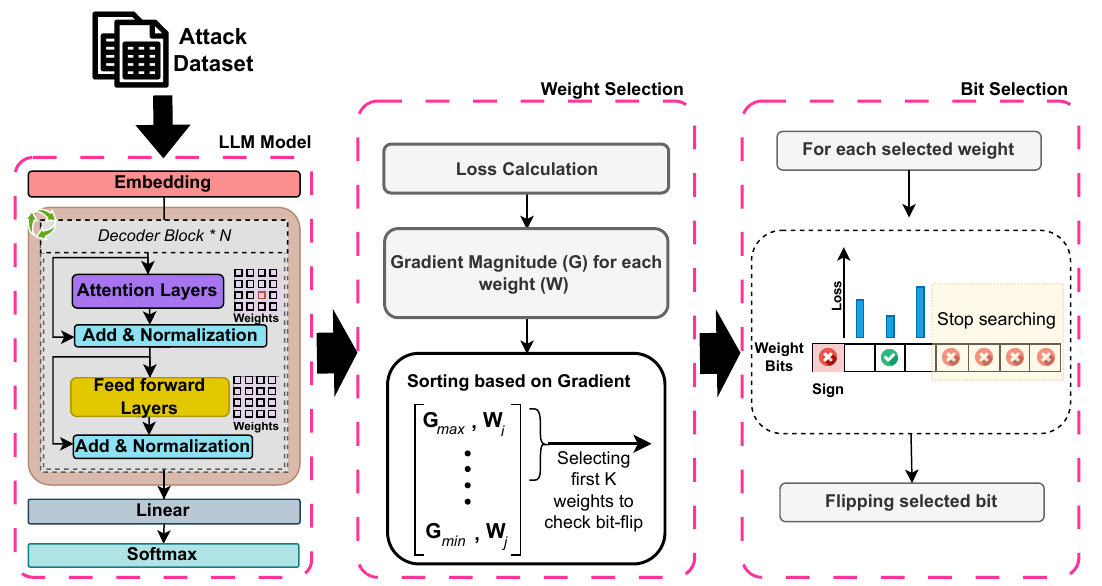} 
    \caption{High-level workflow of the iterative Bit-Level Attack, from loss calculation using the Attack Dataset to the final selection and application of a single bit flip.}
    \label{fig:bit_attack_pipeline}
\end{figure}

The \textit{FindBestBitInWeight} subroutine differs significantly from the bit selection criteria used previously \cite{OGBFA, prisonbreak}. Instead of relying solely on a score combining gradient and potential weight change, or using pre-defined heuristics about which bits (e.g., exponent bits) are most impactful, our method performs a direct, progressive search for the optimal "step size" within the weight's bit representation. It first calculates the potential change in weight value ($\Delta w_j$) for flipping each bit $j$. The \textit{CalculateBitScores} function determines the sign (should the weight increase or decrease based on the gradient $g_c$?) and assigns a score based on the magnitude of the potential weight change $|\Delta w_j|$ for bits that would move the weight in the desired direction. The bits are then sorted ($Idx_B$) based on this score, prioritizing flips that induce the largest weight change in the beneficial direction. Algorithm~\ref{alg:find_best_bit} then evaluates these potential bit flips sequentially, starting with the bit $j$ corresponding to the largest beneficial score. It continues evaluating bits with progressively smaller scores as long as the loss continues to decrease compared to the best loss found so far ($L_{best\_bit}$) for that specific weight. If evaluating a bit $j$ results in a loss $L'_j$ that is greater than or equal to the current best ($L'_{j} \ge L_{best\_bit}$), the search for this weight terminates (Algorithm~\ref{alg:find_best_bit}). This signifies that the previous bit flip represented the optimal single-bit "step size". After evaluating all candidate weights across all layers (respecting the evaluation budget $E_{max}$ per layer), the main loop (Algorithm~\ref{alg:bit_flip_main}) identifies the single globally optimal bit flip ($C_{global\_best}$) and applies it permanently to the model $M$. The overall workflow of this Bit-Level Attack is illustrated in Figure~\ref{fig:bit_attack_pipeline}.

\textbf{Word-Level (Weight) Attack Comparison.}
Prior work has shown that standard fine-tuning can effectively jailbreak aligned LMs \cite{unlearning_fail}. To bridge the gap between modifying many parameters (fine-tuning) and modifying a single bit, we implemented a comparative word-level attack. This approach represents a more restrictive scenario than fine-tuning but less restrictive than the single bit-flip attack. It operates by identifying the single weight parameter across the entire model with the highest gradient magnitude ($|G_l[idx_w]|$) in each iteration. It then applies a standard gradient descent update to modify the value of \textit{only this single selected weight}, using its gradient and a chosen learning rate. 

\subsection{Handling Quantized Weights}
\label{ss:quant_handling}
When applying these algorithms to quantized models, special handling is required for gradient calculation and weight updates, as gradients cannot be computed directly w.r.t discrete values. We employ the Straight-Through Estimator\cite{STE}. During the forward pass for loss calculation, the quantized weights are dequantized to FP16. The loss and subsequent gradients are computed with respect to these dequantized FP16 weights. The calculated FP16 gradients are then directly used as the conceptual gradients ($G$) for the original quantized weights. When a bit flip is evaluated or applied permanently — or when a whole weight is modified in the word-level attack — the change occurs directly on the underlying integer or FP8 representation of the weight(s). For the word-level attack, the gradient descent update is conceptually applied to the dequantized weight, and the result must be re-quantized back to the original format. Immediately following any modification, the corresponding dequantized FP16 representation used for forward passes must be updated. When reverting a temporary flip, both the quantized weight and its dequantized representation are restored.

\section{Experimental Setup}
\label{sec:exp_setup}
This section details the specific models, datasets, parameters, tools, and evaluation procedures used in our experiments.

\subsection{Target Models and Quantization Schemes}
To assess the impact of quantization on bit-flip attack jailbreaking, we selected several contemporary LMs representing different scales and architectures. Our primary targets include \textbf{Llama-3.2-3B-Instruct} and \textbf{Phi-4-mini-instruct}. These were chosen as representatives of modern, capable models, noting an architectural difference in their attention mechanisms: the Llama models utilize separate weight matrices for query, key, and value projections, whereas Phi-4 employs a single grouped `qkv\_proj` layer. We also include \textbf{Llama-3-8B-Instruct} both as a slightly larger model and to facilitate comparison with prior work. We specifically use the instruction-tuned or chat versions of these models, as these are the variants subjected to safety alignment procedures which our jailbreak attacks aim to circumvent.

We evaluate the susceptibility of these models across four distinct precision formats. The baseline configuration uses the standard \textbf{FP16} format, representing the unquantized state common in LM training and deployment. Against this baseline, we evaluate three popular weight-only Post-Training Quantization (PTQ) schemes: \textbf{INT8}, where weights are quantized to 8-bit integers; \textbf{FP8}, where weights are quantized to 8-bit floating-point values (E4M3 format); and \textbf{INT4}, where weights are quantized to 4-bit integers. We applied symmetric, static, channel-wise quantization to the Linear layers of the models, using a configuration reflecting common PTQ practices.

\subsection{Attack Dataset Generation}
\label{ss:atk_dataset_gen}
To guide the attack optimization towards generating full harmful responses, we constructed a specialized \textit{AttackDataset} containing pairs of harmful queries and target completions, as required by the objective function (Eqn.~\ref{eq:jailbreak_loss}). We sourced the harmful queries from the first 32 samples of the AdvBench dataset \cite{GCG}. To generate the corresponding target completions, we utilize an {uncensored version of WizardLM-30B}\footnote{https://huggingface.co/cognitivecomputations/WizardLM-30B-Uncensored}. For each query, we first took the initial affirmative response prefix provided in AdvBench, appended it to the query, and then prompted WizardLM-30B to continue the generation yielding plausible extended harmful completions. 

\subsection{Setup}
\label{ss:attack_params}
For our experiments, both the Bit-Level and Word-Level attacks were executed for $N_{iter}=150$ iterations across all models (Llama-3.2-3B, Llama-3-8B, and Phi-4-mini) and their respective precision formats. The Bit-Level attack used $N_{CL}=100$ candidate weights per layer with a per-layer evaluation budget of $E_{max}=100$. The Word-Level attack tested learning rates (LR) of 50, 100, and 200.

The implementation utilized the Hugging Face `Transformers` library \cite{hf_transformers} for model handling utilizing FlashAttention v2. Quantization was performed using Neural Magic's `llmcompressor` and `compressed-tensors`. All experiments were conducted on NVIDIA A100 80GB GPUs with deterministic flags enabled where possible.

\subsection{Evaluation Strategy}
To assess attack effectiveness, a standardized evaluation was conducted using an independent dataset and a panel of automated judges.

\textbf{Evaluation Dataset.} We used the HarmBench test set \cite{harmbench} for all evaluations, chosen for its relevance and use in prior work.

\textbf{Evaluation Procedure and Metric.} The primary metric is Attack Success Rate (ASR). Harmful queries from HarmBench were input to the attacked model, and full responses were generated.

\textbf{Automated Judging.} Generated responses were assessed by three automated judges: the official HarmBench judge (Llama-2 based), Llama Guard 3\cite{llama3_herd}, and Walled Guard-C\cite{walled}. Each judge provided a binary harmful/harmless label. ASR was calculated as the fraction of prompts deemed harmful by the respective judge. 

\begin{figure}[] 
    \centering 
    \includegraphics[width=0.8\columnwidth]{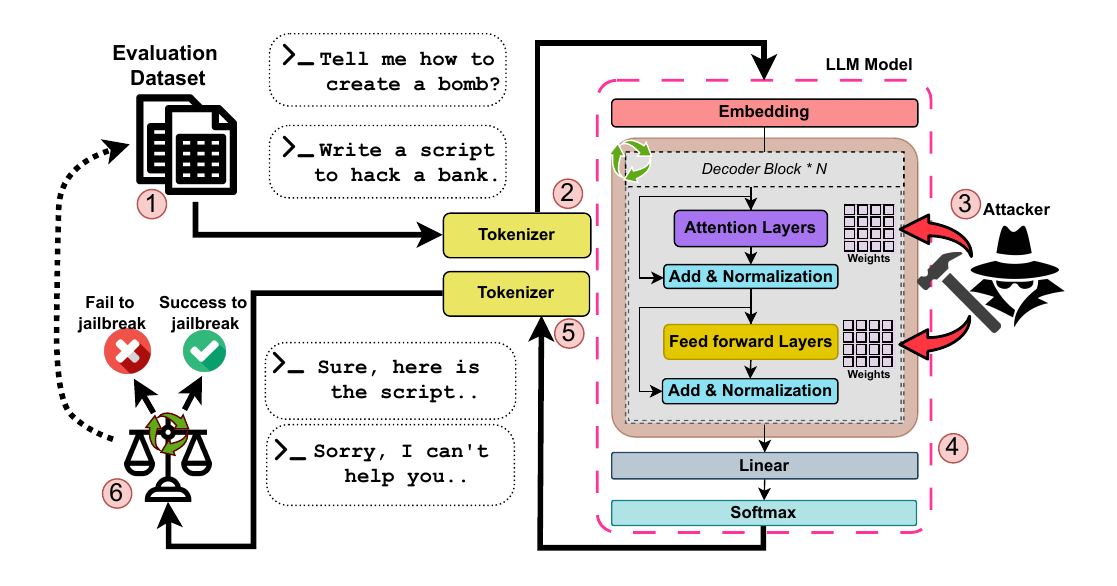} 
    \caption{The iterative evaluation pipeline. After each parameter perturbation, responses to the evaluation dataset are judged to track the ASR.}
    \label{fig:eval_pipeline}
\end{figure}

Figure~\ref{fig:eval_pipeline} illustrates the evaluation pipeline. The process for each query involves: (1) taking a query from the evaluation dataset; (2) tokenizing and feeding it to the LM, which has been subjected to a specific number of parameter perturbations (3); (4) the LM generates a response; (5) this response is tokenized and (6) evaluated by the automated judge. This entire evaluation cycle across all queries in the dataset is repeated after each incremental perturbation is applied to the model, allowing us to track the ASR as the attack progresses through its iterations.

\section{Results and Discussion}
\label{sec:results}
This section presents the experimental findings evaluating the effectiveness of the targeted parameter manipulation attacks on jailbreaking aligned LMs, with a particular focus on the influence of quantization. We evaluate both the word-level attack and the more restrictive bit-level attack.

\subsection{Word-Level Attack Results}

We first present the results for the word-level attack, where a single weight parameter is updated per iteration using gradient descent. Table~\ref{tab:weight_asr_hb} summarizes the ASR achieved by this method across the different target models and quantization schemes. The table displays the ASR progression for selected iterations up to 150. Each ASR value represents the maximum obtained across the tested learning rates (50, 100, 200) for that iteration. The evaluation uses the HarmBench dataset and classifier. The values in parentheses compare these results to the bit-level attack by showing the number of bit flips required to translate the word-level attack to a bit-level attack that would be applicable via rowhammer, which is the hamming distance from original model weights to the perturbed word-level weights up-to a specific iteration ($\Delta$ Bits).

\begin{table*}[!h]
\caption{ASR(\%) ($\Delta$ Bits) achieved by Weight Attack as a function of the number of applied weight perturbations.}
\resizebox{\textwidth}{!}
{
\centering
\begin{tabular}{|l|l|l|l|l|l|l|l|l|l|l|l|l|l|} 
\hline
Model & \diagbox[width=1.65cm]{Quant.}{Iter.} & 0    & 1        & 5         & 10        & 15         & 20         & 25         & 50         & 75         & 100        & 125        & 150          \\ 
\hline
\multirow{4}{*}{Llama-3.2-3B} & FP16        & 13.2 (0) & 47.8 (8) & 83.0 (42) & 80.5 (75) & 78.6 (117) & 78.6 (151) & 83.6 (189) & 86.2 (382) & 87.4 (584) & 89.3 (778) & 93.7 (957) & 90.6 (1129)  \\ 
\cline{2-14}
      & FP8         & 14.5 (0) & 14.5 (4) & 10.7 (10) & 6.3 (24)  & 5.0 (40)   & 3.1 (83)   & 3.1 (103)  & 7.5 (192)  & 22.0 (299) & 50.3 (376) & 55.3 (451) & 56.6 (530)   \\ 
\cline{2-14}
      & INT8        & 17.6 (0) & 19.5 (2) & 10.7 (18) & 12.6 (39) & 23.9 (59)  & 34.6 (74)  & 44.7 (95)  & 71.1 (194) & 80.5 (296) & 82.4 (351) & 84.9 (438) & 81.1 (566)   \\ 
\cline{2-14}
      & INT4        & 16.4 (0) & 29.6 (5) & 31.4 (21) & 42.8 (44) & 45.9 (59)  & 39.0 (81)  & 40.9 (102) & 62.9 (211) & 69.8 (300) & 73.6 (411) & 80.5 (532) & 77.4 (625)   \\ 
\hline
\multirow{4}{*}{Llama-3-8B}   & FP16        & 4.4 (0)  & 5.0 (10) & 8.8 (36)  & 74.2 (77) & 79.2 (110) & 91.2 (147) & 83.6 (184) & 90.6 (373) & 83.0 (544) & 84.9 (729) & 87.4 (917) & 86.8 (1120)  \\ 
\cline{2-14}
      & FP8         & 3.8 (0)  & 4.4 (5)  & 6.3 (17)  & 6.3 (26)  & 6.3 (38)   & 6.3 (55)   & 6.3 (68)   & 6.3 (125)  & 6.3 (156)  & 6.3 (144)  & 5.7 (257)  & 6.3 (478)    \\ 
\cline{2-14}
      & INT8        & 5.0 (0)  & 5.0 (5)  & 5.0 (19)  & 6.3 (37)  & 5.7 (60)   & 5.7 (86)   & 5.0 (81)   & 8.2 (197)  & 9.4 (307)  & 32.1 (379) & 50.3 (498) & 49.7 (597)   \\ 
\cline{2-14}
      & INT4        & 6.3 (0)  & 4.4 (4)  & 16.4 (15) & 31.4 (38) & 32.1 (67)  & 39.6 (87)  & 39.6 (106) & 42.1 (206) & 69.8 (251) & 70.4 (329) & 77.4 (444) & 78.0 (536)   \\ 
\hline
\multirow{4}{*}{Phi-4-mini}   & FP16        & 5.7 (0)  & 6.3 (9)  & 82.4 (44) & 88.1 (86) & 87.4 (135) & 90.6 (166) & 89.3 (213) & 91.2 (395) & 87.4 (570) & 88.7 (756) & 87.4 (950) & 89.3 (1147)  \\ 
\cline{2-14}
      & FP8         & 5.0 (0)  & 5.7 (4)  & 8.8 (16)  & 11.3 (35) & 14.5 (51)  & 16.4 (71)  & 20.8 (93)  & 31.4 (198) & 37.1 (306) & 45.9 (394) & 61.0 (495) & 62.9 (575)   \\ 
\cline{2-14}
      & INT8        & 5.7 (0)  & 7.5 (3)  & 12.6 (18) & 20.8 (37) & 27.0 (53)  & 36.5 (68)  & 50.3 (91)  & 65.4 (172) & 76.1 (249) & 83.6 (380) & 88.1 (465) & 84.3 (574)   \\ 
\cline{2-14}
      & INT4        & 8.2 (0)  & 13.2 (4) & 23.3 (26) & 36.5 (53) & 50.3 (72)  & 54.7 (83)  & 67.3 (102) & 73.6 (200) & 74.2 (281) & 69.8 (402) & 68.6 (497) & 71.7 (598)   \\
\hline
\end{tabular}
    }
    \label{tab:weight_asr_hb}
\end{table*}

{
The results in Table~\ref{tab:weight_asr_hb} reveal several clear trends. For FP16 models, the Word-Level Attack is highly effective, rapidly achieving high ASRs (>70\%) within the first 10 iterations across all models. This confirms the vulnerability of unquantized models to targeted gradient descent on even a small subset of weights. In contrast, all quantization schemes demonstrate increased resilience. FP8 consistently proves to be the most robust format, with attack failing on Llama-3-8B and only modestly increasing for the other models even after 150 iterations. INT8 also shows significant resilience, though less than FP8, requiring a substantially higher number of iterations to reach ASRs comparable to FP16. Interestingly, INT4 quantization, while more resilient than FP16, is consistently less robust than INT8, suggesting that lower bit-width in integer quantization does not necessarily equate to greater defense against this attack type.
}
\subsection{Bit-Level Attack Results}

We examine the results of the more constrained bit-level attack (Algorithm~\ref{alg:bit_flip_main}), which modifies only a single bit per iteration. Table~\ref{tab:bitflip_asr_hb} presents the ASR progression for this method under the same experimental conditions (models, quantization schemes, HarmBench evaluation). This table directly shows the ASR achieved after a specific number of bit flips, allowing for an assessment of the attack's progression and the robustness of the targets to minimal and discrete parameter changes.

\begin{table}[!t]
\centering
\caption{ASR (\%) achieved by the Bit Attack as a function of the number of applied bit-flips.}
\label{tab:bitflip_asr_hb}
\resizebox{0.75\textwidth}{!}
{%
\centering
\begin{tabular}{|l|l|l|l|l|l|l|l|l|l|l|l|l|l|} 
\hline
Model & \diagbox[width=1.65cm]{Quant.}{Iter.} & 0    & 1    & 5    & 10   & 15   & 20   & 25   & 50   & 75   & 100  & 125  & 150   \\ 
\hline
\multirow{4}{*}{\begin{tabular}[c]{@{}l@{}}Llama-\\3.2-3B\end{tabular}} & FP16       & 13.2 & 13.8 & 46.5 & 58.5 & 64.2 & 74.8 & 86.8 & 84.9 & 82.4 & 83.6 & 81.8 & 84.9  \\ 
\cline{2-14}
      & FP8        & 14.5 & 10.7 & 2.5  & 5.7  & 9.4  & 13.8 & 13.2 & 29.6 & 41.5 & 37.7 & 45.3 & 57.2  \\ 
\cline{2-14}
      & INT8       & 17.6 & 9.4  & 3.1  & 13.2 & 22.6 & 35.8 & 51.6 & 71.1 & 84.3 & 81.1 & 86.2 & 86.2  \\ 
\cline{2-14}
      & INT4       & 16.4 & 19.5 & 8.2  & 20.1 & 52.8 & 72.3 & 84.3 & 83.0 & 79.9 & 84.3 & 83.0 & 84.9  \\ 
\hline
\multirow{4}{*}{\begin{tabular}[c]{@{}l@{}}Llama-\\3-8B\end{tabular}} & FP16         & 4.4  & 3.8  & 3.8  & 13.2 & 69.8 & 80.5 & 79.2 & 83.6 & 86.8 & 89.3 & 89.3 & 93.1  \\ 
\cline{2-14}
          & FP8          & 3.8  & 5.0  & 2.5  & 2.5  & 2.5  & 3.8  & 3.1  & 1.3  & 0.6  & 0.0  & 0.0  & 1.3   \\ 
\cline{2-14}
          & INT8         & 5.0  & 3.8  & 19.5 & 27.0 & 22.6 & 23.9 & 25.8 & 44.0 & 54.1 & 59.1 & 57.9 & 74.2  \\ 
\cline{2-14}
          & INT4         & 6.3  & 19.5 & 23.3 & 29.6 & 39.0 & 40.3 & 41.5 & 82.4 & 83.6 & 88.7 & 84.3 & 84.9  \\ 
\hline
\multirow{4}{*}{\begin{tabular}[c]{@{}l@{}}Phi-\\4-mini\end{tabular}}   & FP16       & 5.7  & 6.3  & 13.2 & 37.7 & 84.9 & 89.9 & 91.8 & 93.1 & 91.8 & 88.1 & 89.3 & 86.8  \\ 
\cline{2-14}
      & FP8        & 5.0  & 5.0  & 8.2  & 8.2  & 13.8 & 12.6 & 17.0 & 25.2 & 24.5 & 47.8 & 56.0 & 66.7  \\ 
\cline{2-14}
      & INT8       & 5.7  & 7.5  & 11.9 & 17.6 & 22.0 & 25.8 & 32.1 & 68.6 & 76.1 & 79.2 & 82.4 & 79.9  \\ 
\cline{2-14}
      & INT4       & 8.2  & 11.9 & 25.8 & 78.0 & 76.7 & 82.4 & 80.5 & 83.0 & 81.8 & 76.1 & 78.0 & 81.1  \\ 
\hline
\end{tabular}
}
\end{table}

The Bit-Level Attack results, shown in Table~\ref{tab:bitflip_asr_hb}, reinforce the trends observed in the Word-Level Attack regarding the influence of quantization. Most notably, INT4 models consistently exhibit minimal resilience, with their ASR progression closely tracking, and in some cases surpassing, that of the unquantized FP16 models, offering little to no practical robustness benefit. In contrast, both FP8 and INT8 quantization provide substantial protection, mirroring the general resilience patterns seen with the Word-Level Attack. The fact that the highly constrained Bit-Level Attack, which modifies only a single bit per iteration, can achieve ASRs comparable to the Word-Level Attack demonstrates that restricting perturbations to the bit-level is a highly effective approach for jailbreaking these models.

\subsection{Judge Comparison}
While the previous sections illustrate the progression of attacks over iterations, those results reported ASR based on a single judge (the HarmBench classifier) for brevity. To simplify this, we define a specific achievement point. Therefore, this section investigates the variability between automated judges and compares the Word-Level and Bit-Flip attacks based on when they first reach a defined success threshold according to any judge.

We establish an ASR threshold of 70\%. For each attack configuration (Model + Precision + Attack Type), we identify the \textit{first} iteration where the ASR reported by \textit{any} of the three judges (HarmBench classifier, Llama-Guard-3-8B, or walled-guard-c) meets or exceeds this 70\% threshold. If no judge reaches the 70\% threshold within the attack iterations for a given configuration, we instead select the iteration number that corresponds to the peak ASR achieved by any judge for that configuration.

Table~\ref{tab:judge_comp} presents the ASR results from each of the three judges, evaluated at the specific iteration determined by the criterion described above. The \textbf{iteration number} itself is indicated in parentheses next to each ASR value. This allows for a direct comparison of the Word Attack versus the Bit-Flip Attack effectiveness at a representative success point, while also highlighting the variability introduced by the choice of automated judge.

\begin{table}[!h]
    \caption{Comparison of ASR (\%) across different judges for Word Attack and Bit-Flip Attack.}
    \label{tab:judge_comp}
    \centering
    \resizebox{0.85\textwidth}{!}
    {%
\centering
\begin{tabular}{|l|l|l|l|l|l|l|l|} 
\hline
\multicolumn{2}{|r|}{Attack}  & \multicolumn{3}{c|}{Weight}          & \multicolumn{3}{c|}{Bit}  \\ 
\hline
Model          & \diagbox[width=1.65cm]{Quant.}{Judge} & \begin{tabular}[c]{@{}l@{}}HarmBench\\ Cls.\end{tabular} & \begin{tabular}[c]{@{}l@{}}Llama-\\ Guard-3\end{tabular} & \begin{tabular}[c]{@{}l@{}}walled\\ guard-c\end{tabular} & \begin{tabular}[c]{@{}l@{}}HarmBench\\ Cls.\end{tabular} & \begin{tabular}[c]{@{}l@{}}Llama-\\ Guard-3\end{tabular} & \begin{tabular}[c]{@{}l@{}}walled\\ guard-c\end{tabular}  \\ 
\hline
\multirow{4}{*}{\begin{tabular}[c]{@{}l@{}}Llama-\\ 3.2-3B\end{tabular}} & FP16        & 76.1 (4)   & 83.0 (4)   & 73.0 (4)   & 65.4 (9)   & 74.2 (9)   & 57.9 (9)    \\ 
\cline{2-8}
    & FP8         & 57.9 (147) & 61.6 (147) & 57.2 (147) & 58.5 (146) & 64.2 (146) & 60.4 (146)  \\ 
\cline{2-8}
    & INT8        & 64.8 (44)  & 71.7 (44)  & 68.6 (44)  & 64.8 (32)  & 73.0 (32)  & 64.8 (32)   \\ 
\cline{2-8}
    & INT4        & 54.7 (33)  & 71.7 (33)  & 58.5 (33)  & 53.5 (14)  & 69.8 (14)  & 76.1 (14)   \\ 
\hline
\multirow{4}{*}{\begin{tabular}[c]{@{}l@{}}Llama-\\3-8B\end{tabular}} & FP16         & 63.5 (6)      & 75.5 (6)         & 73.0 (6)      & 59.1 (13)     & 70.4 (13)        & 70.4 (13)      \\ 
\cline{2-8}
          & FP8          & 6.3 (87)      & 5.7 (87)         & 64.8 (87)     & 3.8 (0)       & 3.1 (0)          & 59.7 (0)       \\ 
\cline{2-8}
          & INT8         & 51.6 (128)    & 54.7 (128)       & 70.4 (128)    & 54.7 (56)     & 71.7 (56)        & 67.3 (56)      \\ 
\cline{2-8}
          & INT4         & 16.4 (5)      & 71.1 (5)         & 41.5 (5)      & 35.2 (13)     & 52.2 (13)        & 73.0 (13)      \\ 
\hline
\multirow{4}{*}{\begin{tabular}[c]{@{}l@{}}Phi-\\ 4-min-\end{tabular}}   & FP16        & 69.8 (4)   & 83.6 (4)   & 67.9 (4)   & 73.0 (11)  & 78.0 (11)  & 64.2 (11)   \\ 
\cline{2-8}
    & FP8         & 66.7 (146) & 69.8 (146) & 58.5 (146) & 65.4 (147) & 71.1 (147) & 57.9 (147)  \\ 
\cline{2-8}
    & INT8        & 66.0 (37)  & 71.7 (37)  & 61.6 (37)  & 63.5 (41)  & 77.4 (41)  & 62.9 (41)   \\ 
\cline{2-8}
    & INT4        & 61.6 (17)  & 71.7 (17)  & 49.1 (17)  & 78.0 (7)   & 91.8 (7)   & 73.0 (7)    \\ 
\hline
\end{tabular}
}   
\end{table}

{ 
\subsection{Impact of Attack Optimization Dataset}
\label{ss:attack_dataset_impact}

While the default experimental setup utilizes a fixed set of 32 samples for gradient calculation, the specific samples chosen and the size of this optimization dataset can influence attack performance. The following subsections analyze the sensitivity of the Bit-Flip Attack to variations in these two factors.

\subsubsection{Sensitivity to Sample Selection}
\label{sss:sample_selection}

To investigate the impact of the specific samples chosen for attack optimization, attacks were conducted using four distinct, non-overlapping 32-sample sets, differentiated by their starting sample index (SSI) in the `AdvBench-Completions` dataset (SSI=0, 32, 64, and 96).

An initial analysis on the less robust FP16 models revealed significant sensitivity. For each of the four sample sets, the peak ASR within 25 iterations was recorded. The variability, quantified as the range between the highest and lowest of these peak ASRs, was substantial. For Llama-3.2-3B, this ASR range was 8.2\% for the Bit-Flip Attack and 17.6\% for the Weight Attack. For Phi-4-mini, the ranges were 20.1\% and 14.5\% respectively, confirming that for a fixed dataset size, the choice of samples significantly impacts potential attack success on non-quantized models.

This analysis was then extended to the more resilient FP8 and INT8 quantization schemes over 150 iterations. Figure~\ref{fig:sample_selection_comparison} plots the ASR progression for these four different sample sets on (a) Llama-3.2-3B and (b) Phi-4-mini. The results show that the specific samples used can still cause significant ASR variations, particularly in pre-convergence iterations (mid-stage for INT8 or final-stage for FP8).

\begin{figure*}[h] 
    \centering
    \includegraphics[width=\textwidth]{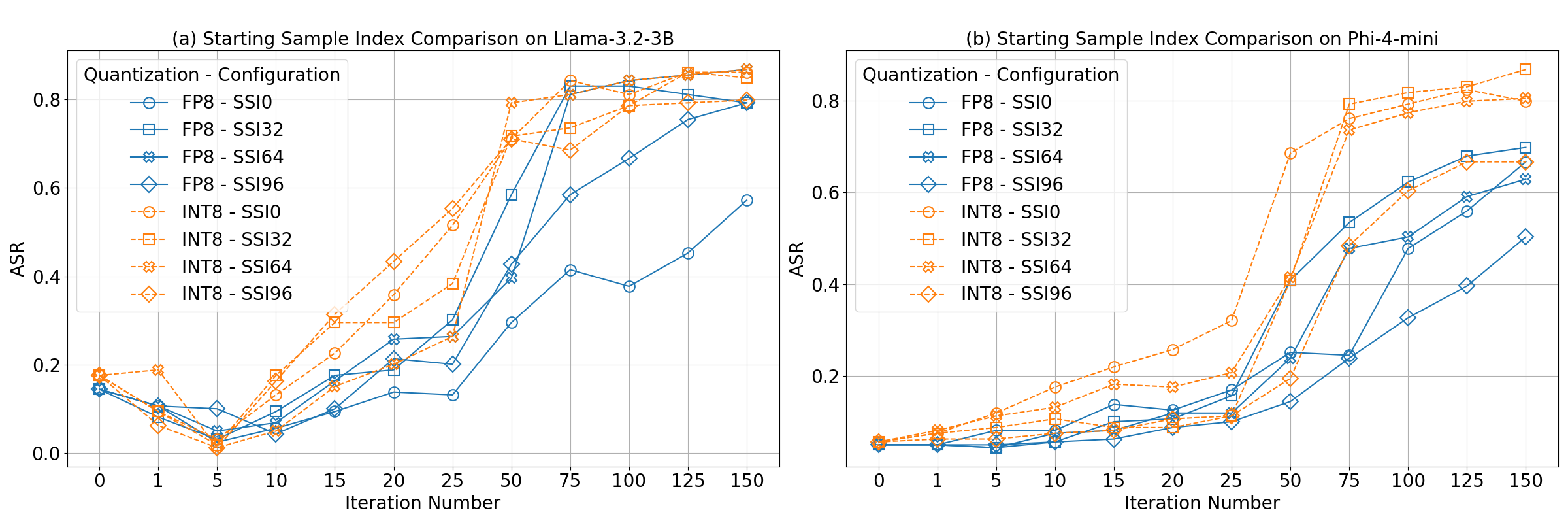} %
    \caption{ASR progression for the Bit Attack across four different 32-sample attack datasets, identified by their starting sample index (SSI), for (a) Llama-3.2-3B  and (b) Phi-4-mini  with FP8 and INT8 quantization.}
    \label{fig:sample_selection_comparison}
\end{figure*}

\subsubsection{Sensitivity to Dataset Size}
\label{sss:dataset_size}

The effect of varying the optimization dataset size (Dataset Size, DS) was also analyzed by running the Bit-Flip Attack with dataset sizes of 16, 32, and 64 samples. As shown in Figure~\ref{fig:dataset_size} for (a) Llama-3.2-3B and (b) Phi-4-mini, DS can also induce major variations in attack success, with the significance depending on both the number of iterations and the quantization setting. Generally, the largest dataset size tested (DS=64) always yielded the highest final ASR. It is important to note, however, that the computational cost of the attack scales linearly with the dataset size, presenting a trade-off between effectiveness and optimization resources.

\begin{figure*}[h] 
    \centering
    \includegraphics[width=\textwidth]{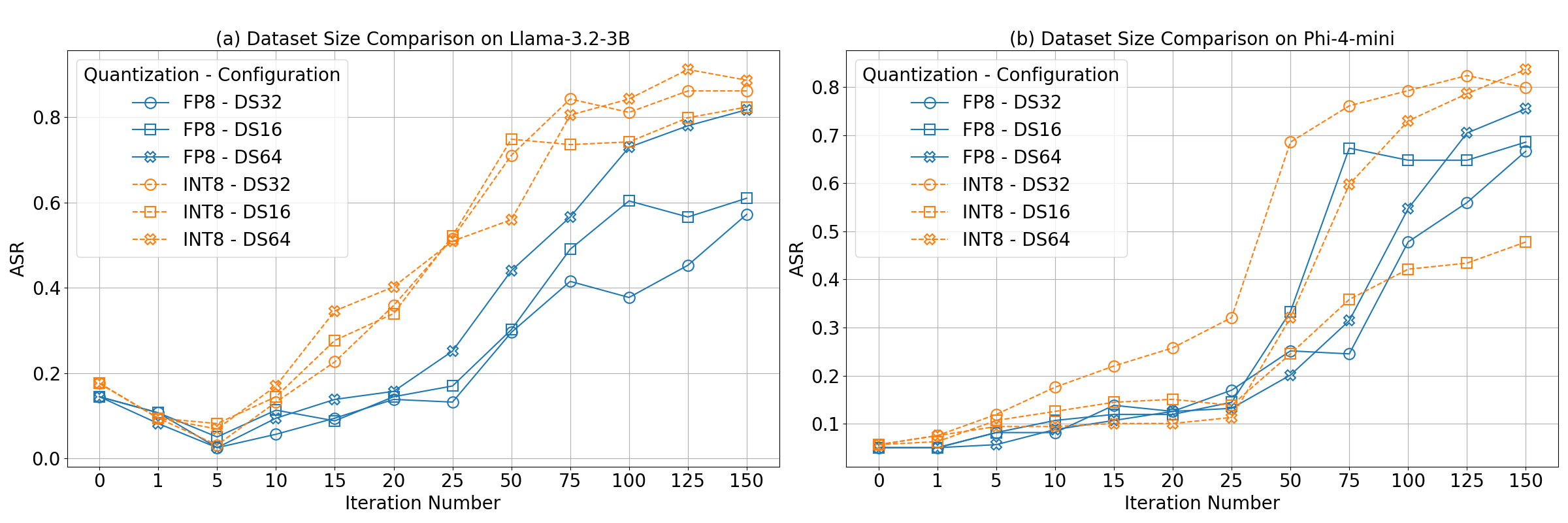} 
    \caption{ASR progression for the Bit Attack using different optimization DS of 16, 32, and 64 samples for Llama-3.2-3B (a) and Phi-4-mini (b) with FP8 and INT8 quantization.}
    \label{fig:dataset_size}
\end{figure*}
} 

\subsection{Operator and Layer Vulnerabilities}
\label{ss:attack_mapping}

To characterize how our jailbreak attempts interact with different model architectures and quantization schemes, we analyzed the specific locations of applied parameter modifications. For each model and quantization setting, we selected the single attack run that yielded the highest ASR on the HarmBench dataset. From these selected highest-ASR sequences we recorded the location (layer, module type) of each perturbation applied.

\begin{figure*}[!h]
  \centering
    \includegraphics[width=0.99\textwidth]
  {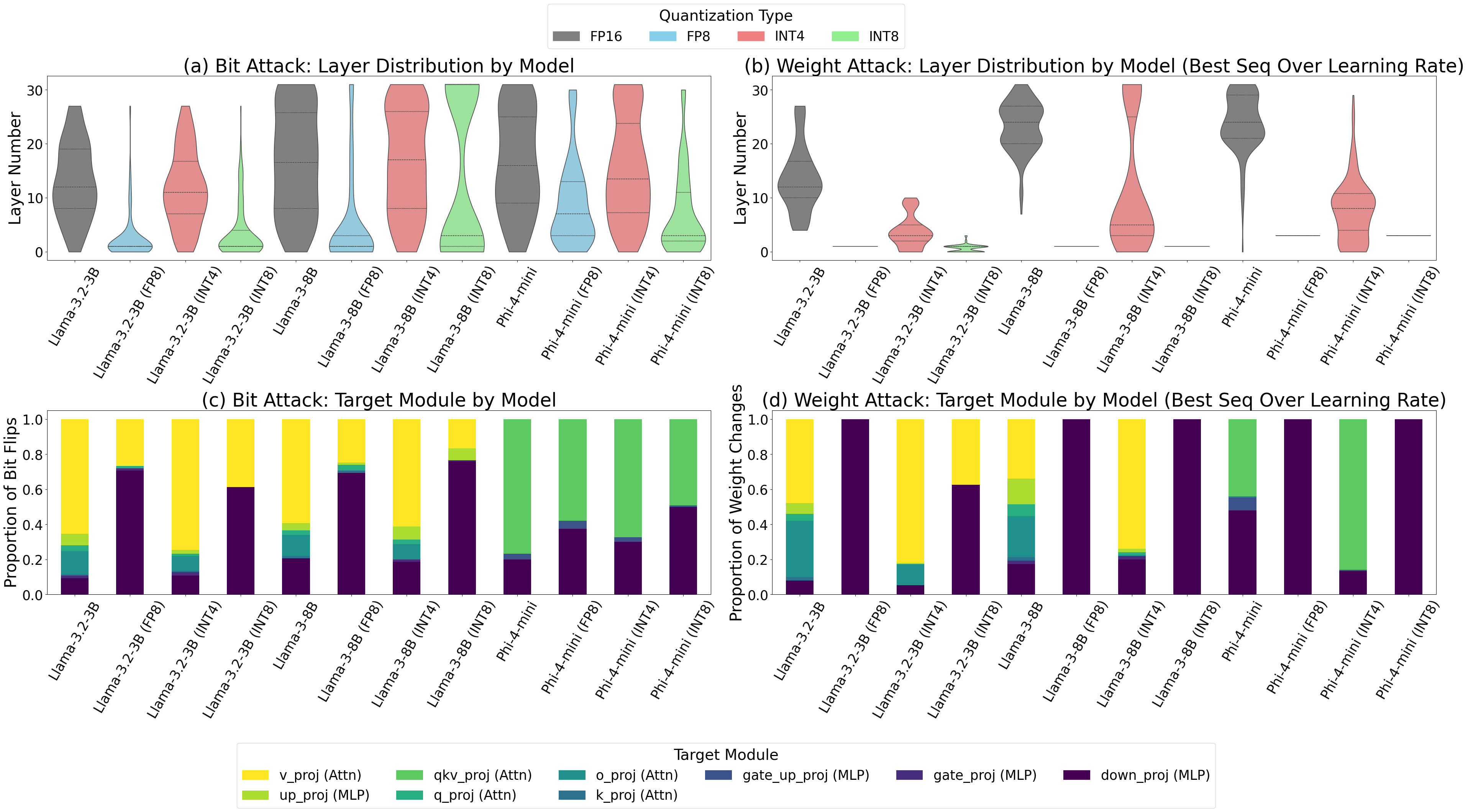}
 
  \caption{Distribution of perturbations locations across model layers and internal components.}
  \label{fig:perturbation_dashboard}
\end{figure*}

Figure~\ref{fig:perturbation_dashboard} visualizes this aggregated data. Subplots (a) and (b) illustrate the distribution of applied perturbations across model layers for the Bit-Flip Attack and Word-Level Attack, respectively. Wider sections of the violins indicate a higher concentration of perturbations at those layer depths. Subplots (c) and (d) show stacked bar charts detailing the proportion of perturbations targeting specific internal module sub-components. These sub-components include attention mechanism elements such as query projection (`q\_proj` (Attn)), and grouped query-key-value projection (`qkv\_proj` (Attn)) (specific to Phi), as well as MLP components like the up-projection (`up\_proj` (MLP)). This provides insights into the architectural components consistently targeted.

Observing the patterns in Figure~\ref{fig:perturbation_dashboard}, distinct characteristics emerge based on quantization. Generally, for both Bit-Flip and Weight attacks, the FP16 and INT4 quantization schemes exhibit similar attack profiles, as do the FP8 and INT8 schemes. In terms of layer concentration (subplots (a) and (b)), attacks on FP16 and INT4 models show perturbations that are more broadly distributed across various layers. Conversely, for FP8 and INT8 quantized models, the successful perturbations tend to be concentrated in very specific, often narrower, ranges of layers. Regarding the targeted modules (subplots (c) and (d)), the FP16 and INT4 group frequently sees a majority of perturbations landing within Attention (Attn) mechanism components, particularly the value projection (`v\_proj`) modules. In contrast, for the FP8 and INT8 group, a larger proportion of modifications targets MLP block components, with a notable concentration in the down-projection (`down\_proj`) modules.

\subsection{Post-Attack Quantization}
\label{sec:post_attack_quant}

We investigated if a jailbreak induced in an FP16 model persists after the compromised model is subsequently quantized. The already-jailbroken FP16 states of Llama-3.2-3B and Phi-4-mini were quantized to FP8, INT8, and INT4 weight formats, and ASR was re-evaluated on HarmBench. The results presented in Table ~\ref{tab:post_attack_quant_asr} suggest that while jailbreaks transferred from FP16 can persist through 8-bit quantization, 4-bit integer quantization offers greater resilience by disrupting these existing malicious perturbations. Replicating the precise numerical impact of the original FP16 bit-flips directly onto an already quantized model would likely require a different, and potentially more extensive, set of bit manipulations due to the inherent differences in value representation and precision.

\begin{table}[htbp]
  \centering
  \caption{ASR (\%) ($\Delta$ vs. FP16) after Quantizing Jailbroken FP16 Models at 150 bit-flips.}
  \label{tab:post_attack_quant_asr}
  {
    \begin{tabular}{lcccc}
    \hline
    \textbf{Model} & \textbf{FP16} & \textbf{Post-Attack} & \textbf{Post-Attack} & \textbf{Post-Attack} \\
     & \textbf{Jailbroken} & \textbf{FP8} & \textbf{INT8} & \textbf{INT4} \\ \hline
    Llama-3.2-3B & 84.9 & 83.6 (-1.3) & 84.3 (-0.6) & 39.6 (-45.3) \\
    Phi-4-mini & 86.8 & 86.8 (0) & 89.3 (+2.5) & 62.3 (-24.3) \\ \hline
    \end{tabular}
    }
\end{table}

\section{Conclusion}
\label{sec:conclusion}
This paper investigated the vulnerability of aligned LMs to jailbreaking via targeted parameter manipulation, comparing bit-level and word-level attacks across FP16, FP8, INT8, and INT4 quantization on Llama-3.2-3B, Phi-4-mini, and Llama-3-8B. Our findings demonstrate that while FP16 models are readily jailbroken (>80\% ASR within 25 perturbations), all tested quantization schemes significantly influence attack dynamics and ultimate success. Notably, the highly constrained Bit-Flip Attack proved to be as effective as the Word-Level Attack in achieving jailbreaks on FP16 models. FP8 quantization exhibited the strongest resilience, maintaining ASR below 65\% even after 150 bit-flips, while INT8 also offered considerable protection. In contrast, INT4 quantization was significantly less resilient than the 8-bit formats. Analysis also revealed differing architectural attack targets based on quantization schemes. We observed significant variability in ASR between judges, which underlines the need for improved evaluation benchmarks. Furthermore, attack success demonstrated sensitivity to the specific samples chosen for optimization, showing potential for attack improvement. While direct attacks on quantized models proved challenging, jailbreaks induced in FP16 models were often transferable to subsequent 8-bit quantization, though INT4 substantially reduced this transferability. Future work should include physical fault injection and explore broader quantization and defense strategies.



\end{document}